\documentclass[aps,twocolumn,prb]{revtex4}
\usepackage{graphicx}
\usepackage{amssymb}

\begin{document}

\title{Fluctuation Study of the Specific Heat of Mg$^{11}$B$_{2}$ }
\author{Tuson Park}
\author{M. B. Salamon}
\affiliation{Department of Physics and Material Research Laboratory, University of\\
Illinois at Urbana-Champaign, IL 61801, USA}
\author{C. U. Jung}
\author{Min-Seok Park}
\author{Kyunghee Kim}
\author{Sung-Ik Lee}
\affiliation{National Creative Research Initiative Center for Superconductivity and
Department of Physics, Pohang University of Science and Technology, Pohang\\
790-784, Republic of Korea}
\date{Revised June 13th, 2002}

\begin{abstract}
The specific heat of polycrystalline Mg$^{11}$B$_{2}$ has been measured with
high resolution ac calorimetry from 5 to 45 K at constant magnetic fields.
The excess specific heat above T$_{c}$ is discussed in terms of Gaussian
fluctuations and suggests that Mg$^{11}$B$_{2}$ is a bulk superconductor
with Ginzburg-Landau coherence length $\xi _{0}=26$ \AA . The
transition-width broadening in field is treated in terms of
lowest-Landau-level (LLL) fluctuations. That analysis requires that $\xi
_{0}=20$ \AA . The underestimate of the coherence length in field, along
with deviations from 3D LLL predictions, suggest that there is an influence
from the anisotropy of B$_{c2}$ between the c-axis and the a-b plane.
\end{abstract}

\maketitle

Experimental observations of thermodynamic fluctuations in the specific heat
have been limited in low-T$_{c}$ superconductors because the long coherence
lengths make the excess specific heat very small compared to the mean-field
term \cite{tinkham}. By contrast, the high transition temperatures and small
coherence lengths of cuprate superconductors lead to significant fluctuation
effects \cite{salamon}. In the recently discovered superconductor Mg$^{11}$B$%
_{2}$ \cite{Akimatsu}, the coherence length and superconducting transition
temperature lie between these extremes, suggesting that fluctuation effects
will be observable and lead to further information on the superconducting
coherence length. Indeed, the excess magnetoconductance of Mg$^{11}$B$_{2}$
was reported recently and discussed in terms of fluctuation effects \cite%
{Kang}. Here we report the specific heat of Mg$^{11}$B$_{2}$ from 5 K to 45
K at several magnetic fields. Using high resolution ac calorimetry, we could
study the superconducting transition region in detail. At zero-field, the
excess specific heat is treated in terms of 3D Gaussian fluctuations and in
field, the broadening and shift of the transition is analyzed in terms of
lowest-Landau-level (LLL) fluctuations.

Polycrystalline Mg$^{11}$B$_{2}$ was prepared at $T=950$ C and $p=3$ GPa
from a stoichiometric mixture of Mg and $^{11}$B isotope using a
high-pressure synthesis method. Since the sample was synthesized at high pressure, there has been no additional annealing. Details of the synthesis can be found
elsewhere \cite{jung}$^{,}$\cite{Jung2}.

\begin{figure}
\centering 
\includegraphics[width=8cm,clip]{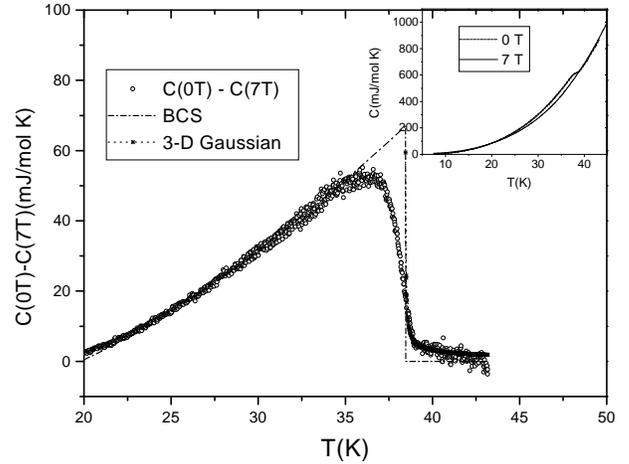}
\caption{Temperature dependence of $\Delta C$ at zero field. The dash-dotted
line is a\ BCS fit with $\Delta C_{\exp }$/$\protect\gamma _{n}T_{c}=0.7$.
The star represents a 3D Gaussian fluctuation model above T$_{c}$
of 38.4 K. Inset: The temperature dependence of the specific heat at 0 and 7
T from 5 to 45 K. }
\label{figure1}
\end{figure}
Measurements of the heat capacity were based on an ac-calorimetric technique 
\cite{ac-method}. A long cylindrical sample was cut into a disk by a diamond
saw and then was sanded to a thin rectangular shape whose dimensions are
approximately $1.1\times 1.5\times 0.1$ mm$^{3}$; its mass is $375$ $\mu $g.
The front face of the prepared sample was coated with colloidal graphite suspension (DAG) thinned with isopropyl alcohol to prevent a possible change of the optical absorption properties of the sample with temperature. The sample was weakly coupled to
the heat bath through helium gas and suspending thermocouple wires. As a
heating source, we used square-wave modulated laser pulse. The oscillating
heat input incurred a steady temperature offset (or dc offset) from the heat
bath with an oscillating temperature superposed. The ac part was kept less
than 1/10 of the dc offset and was then converted to heat capacity by the
relationship : $C\varpropto 1/T_{ac}$. The heat capacity obtained was
converted to a specific heat by using a literature value above the
superconducting temperature \cite{Bouquet}. The frequency of the periodic
heating was chosen so that the $ac$ temperature was inversely proportional
to the frequency and, therefore, to the heat capacity; $23$ Hz was used in
this experiment. The $ac$ and $dc$ temperatures were measured by type E thermocouple, which were varnished on the back face of the sample using a minute amount of GE7031 diluted with a solvent of methanol and toluene. The GE varnish typically amounts to less than 1 \% of the sample mass. Since the field induced error of type E thermocouple is less than 1 \% at 40 K in 8 tesla, we will neglect the field dependence of the addenda contribution (DAG, GE-varnish, and type E thermocouple) and treat the field dependence as due only to the sample.

The inset in Fig.1 shows the temperature dependence of the specific heat at
zero and 7 tesla from 5 to 45 K. The main graph is a plot of $\Delta C_{\exp
}$ vs temperature at zero field. Here $\Delta C_{\exp }$ is the measured
difference between the mixed- and normal-state specific heats. A 7-Tesla
data set was used as a reference state above 20 K because it shows no
observable transition in that range. The subtraction was executed without
any smoothing of the 7-T data. The dash-dotted line is a BCS fit with the
ratio of $\Delta C_{\exp }$/$\gamma _{n}T_{c}$ being variable \cite{Muschel}%
. The normal electronic coefficient $\gamma _{n}$ was set to be 2.6 mJ/mol K
from the literature \cite{Bouquet} and T$_{c}$ of 38.4 K was determined from
scaling discussed below. The best fit showed that the ratio is 0.7, which is much smaller than the weak coupling BCS value of 1.43. Since the ratio is
generally larger for strong coupling superconductors, the small value does
not tell us anything about its coupling strength. Recently, there has been a plethora of experimental and theoretical evidence which supports two-gap features in MgB2, which can explain the non-BCS jump magnitude with some success. \cite{two-gap,wang,yang,liu,choi} However, we cannot rule out such other scenarios as an anisotropic gap structure. \cite{anisotropic-gap3}$^{,}$\cite{anisotropic-gap2} For a system in which fluctuation effects are pronounced, the experimentally determined transition temperature is lower
than the mean-field critical temperature ($T_{c}^{mf}$) because fluctuations
drive the system into the normal state even below $T_{c}^{mf}$. It is
unlikely, however, that this can explain the large deviation from the BCS
value.

\begin{figure}
\centering 
\includegraphics[width=8cm,clip]{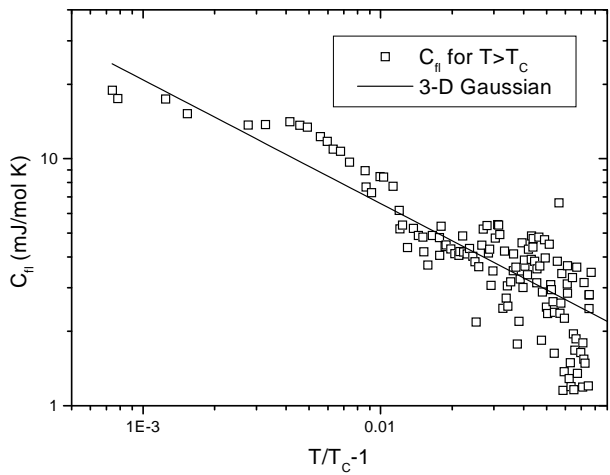}
\caption{ The excess specific heat, C$_{fl}=\Delta C_{\exp }(0),$ for T ${>}$ T$_{c}$ is plotted against reduced temperature $%
t=T/T_{c}-1$ on a log-log scale. The solid line describes Gaussian
fluctuations for a bulk superconductor with a G-L coherence length of 26 \AA.}
\label{figure2}
\end{figure}
Above the transition temperature, there is an excess specific heat tail
apparent in Fig. 1. Thouless \cite{Thouless1} and subsequently Aslamazov and
Larkin \cite{aslamazov} showed that Gaussian fluctuations arise above T$_{c}$
and predicted that $C_{fl}=C^{+}t^{-(2-d/2)}$ with $C^{+}=(k_{B}/8\pi )\xi
_{GL}(0)^{-3}$, where $t=T/T_{c}-1,$ $d$ is the dimensionality, and $\xi
_{GL}(0),$ the T=0 K Ginzburg-Landau coherence length. Figure 2 shows the
temperature dependence of the excess specific heat on a log-log scale. The
data follow a power law with an exponent of -0.5 and C$^{+}=0.66$ mJ/mol K.
The exponent indicates that Mg$^{11}$B$_{2}$ is a 3D superconductor and the
substitution of C$^{+}$ into the above formula gives $\xi _{GL}(0)=26 \pm 1$ \AA .

\begin{figure}
\centering 
\includegraphics[width=8cm,clip]{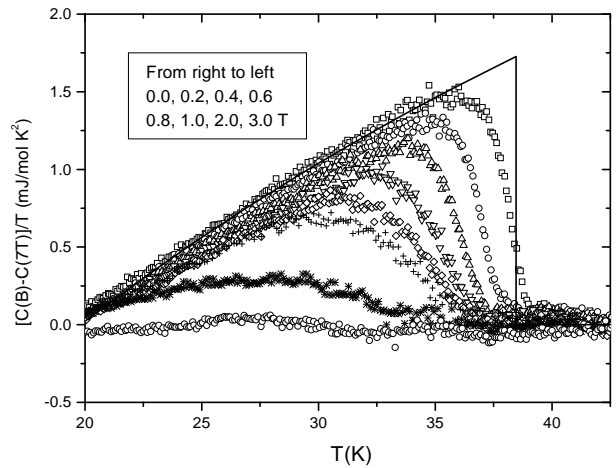}
\caption{$\Delta C/T$ is plotted against temperature in B=0, 0.2, 0.4, 0.6,
0.8, 1.0, 2.0, and 3.0 tesla. Data were taken with increasing temperature after
field cooling. The solid line represents a BCS fit. }
\label{figure3}
\end{figure}
When a magnetic field is applied, the specific heat broadens. Figure 3 shows
the temperature dependence of $\Delta C/T$ at several magnetic fields. The
ratio of the transition temperature shift to the transition width broadening
in field is unique in that it is not as large as in low-T$_{c}$
superconductors nor as small as in high-T$_{c}$ materials. Its intermediate
behavior is related to the fact that the coherence length and the
superconducting transition temperature of Mg$^{11}$B$_{2}$ are intermediate
between low-T$_{c}$ and high-T$_{c}$ superconductors. Lee and Shenoy \cite%
{lee} studied fluctuation phenomena in the presence of a magnetic field,
arguing that bulk superconductors exhibit a field-induced effective change
to one-dimensional behavior in the vicinity of the transition temperature $%
T_{c}(B)$. In a uniform magnetic field, the fluctuating Cooper pairs move in
quantized Landau orbits and, close to upper critical field (B$_{c2}$), the
lowest Landau level dominates the contribution to the excess specific heat.
So, a bulk superconductor behaves like an array of one-dimensional rods
parallel to the field. Thouless \cite{thouless2} extended the idea above and
below $T_{c}$ and suggested a scaling parameter for the fluctuation specific
heat that is valid throughout the transition region : 
\begin{equation}
\frac{\Delta C_{\exp }}{\Delta C_{mft}}=g\left( \frac{t}{\tau }\right) ,
\end{equation}%
\newline
where $t$ is the reduced temperature and $\tau $ is a field dependent
dimensionless parameter that describes the superconducting transition width.
The functional form $g(y)$ is model dependent. When a Hartree-like
approximation \cite{bray} is used to examine the fluctuation effects of the
quartic term in the free energy functional, it results in a simple form : 
\begin{equation}
g(y)=\left( 1+x(y)\right) ^{-1},
\end{equation}%
\begin{equation}
y=x^{2/3}(1-2/x),
\end{equation}%
where $y=t/\tau (B)$.

\begin{figure}
\centering \includegraphics[width=8cm,clip]{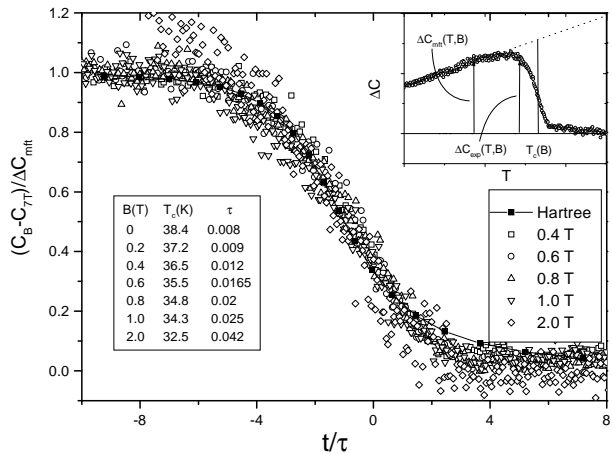}
\caption{$\Delta C_{\exp }/\Delta C_{mft}$ vs $t/\protect\tau $ is shown, where $%
\Delta C_{\exp }=C(B)-C(7T)$ and $\Delta C_{mft}$ are defined in the inset. Here $t$ is reduced temperature and $\protect\tau $ is a field dependent dimensionless broadening parameter. Solid squares are from a Hartree-like approximation. Other field data were scaled so that they were collapsed onto the Hartree result. The obtained parameters (T$_{c}$, $\protect\tau $) are tabulated inside the figure at each magnetic field. The inset sketches the definition of $\Delta C_{mft}(T,B)$ \cite{Farrant-Gough}}
\label{figure4}
\end{figure}
In Fig. 4, the ratio of $\Delta C_{\exp }/\Delta C_{mft}$ in the transition
region was plotted as $t/\tau (B)$, where $\Delta C_{\exp }(B)=C(B)-C(7T)$
and $\Delta C_{mft}$ was determined as in the classic work by Farrant and Gough \cite{Farrant-Gough} by fitting the low temperature side of $\Delta C_{\exp }(B)$ as in Figs 1 and 3 and extrapolating linearly above T$_{c}(B)$. A sketch is shown in the inset of Fig. 4. The scaling parameters $\tau (B)$ and T$_{c}$ were chosen to make the data collapse onto the Hartree-like approximation (solid-squares). The values of $\tau (B)$ and T$_{c}$ are listed in Fig. 4. The temperature dependence of the upper critical field T$_{c}$(B) is plotted in Fig. 5, and shows positive curvature close to $T_{c}(B=0)$. A simple empirical formula \cite{sanchez}, $B_{c2}(T)=B_{c2}(0)[1-(T/T_{c})^{2}][1-a(T/T_{c})^{2}]$, was used to
describe the curvature, in which $a$ is a fitting parameter that is 0 and
0.3 for two-fluid model and for WHH model \cite{whh} respectively. The best
fit, solid line in Fig. 5, was produced with \ $B_{c2}(0)=15.4$ $tesla$ and 
$a=0.8$. Positive curvature near $T_{c}(B=0)$ was also observed in
non-magnetic rare-earth nickel borocarbides $RNi_{2}B_{2}C(R=Lu,Y)$ and
could be explained by the dispersion of the Fermi velocity using an
effective two-band model \cite{shulga}.

\begin{figure}
\centering \includegraphics[width=8cm,clip]{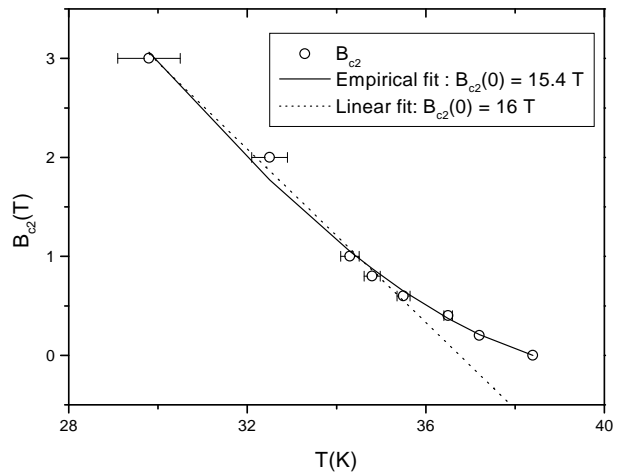}
\caption{ The temperature dependence of the upper critical field B$_{c2}(T)$
is shown. The solid line is a simple empirical formula, $%
B_{c2}(0)[1-(T/T_{c})^{2}][1-a(T/T_{c})^{2}]$, with $a=0.8$ and $%
B_{c2}(0)=15.4$ T. The dotted line is a linear fit with $B_{c2}(0)=16$
T and $dB_{c2}/dT=-0.44$ $T/K.$}
\label{figure5}
\end{figure}
The broadening parameter $\tau $(B) consists of a field dependent part ($%
\tau _{B})$ and a field independent part ($\tau _{in})$. We postulate that
they are independent of each other and add in quadrature, such that the
total broadening is $\tau ^{2}=\tau _{in}^{2}+\tau _{B}^{2}$. The field independent part was obtained in zero field and accounts for sample inhomogeneity and zero-field fluctuation effects while the field dependent part is due solely to field-induced fluctuations. The field dependence of the broadening parameter is given by \cite{thouless2} : 
\begin{equation}
\tau _{B}=\left( \frac{B}{B_{W}}\right) ^{1/\alpha },
\end{equation}%
\begin{equation}
B_{W}=\left( \frac{\Delta C}{k_{B}/8\pi \xi _{0}^{3}}\right) B_{S},
\end{equation}%
where $\alpha =2-(d-2)/2$ ($\alpha =3/2$ for a bulk superconductor) and $%
B_{S}=-T_{c}(dB_{c2}/dT)_{T_{c}}$ in the mean-field scheme. The exponent $d-2
$ indicates a dimensional crossover from d-dimension to d--2 behavior. The
shift field $B_{S}$ is a characteristic field that sets the scale of the
shift of the transition temperature while $B_{W}$ sets the scale of the
width broadening of the transition region. In a standard superconductor, the
ratio $B_{W}/B_{S}$ is very large ($\sim 10^{4}$), and the transition is
shifted much more rapidly in field than it is broadened. In high temperature
superconductors, such as YBCO, the broadening is as large as the shift ($%
B_{S}\sim B_{W})$, which is an indication that a mean-field approach based
on a perturbation expansion might not be proper and that fluctuations should
be treated in the context of critical phenomena. In Mg$^{11}$B$_{2}$, the
ratio is in the order of 10$^{2}$, which is in between those two extremes.
This feature seems consistent with other properties that show aspects of
both conventional and high T$_{c}$ superconductors. In order to study the
anomalous broadening in field, we plot $\tau _{B}$ vs $(B/\Delta C)$ on a
log-log scale in Fig. 6. The slope represents the exponent (1/$\alpha $)
while the coefficient of the slope $\beta $ is related to the
Ginzburg-Landau coherence length through $\beta =(k_{B}/8\pi \xi
_{0}^{3}B_{S})^{1/\alpha }$. The lowest-Landau-level approximation is shown
as a solid line having a slope of 2/3 and coefficient 2.7 $($A/m K$)^{2/3}$. From the coefficient of the fit, the Ginzburg-Landau coherence length is
estimated to be 20 \AA . In the above analysis, the shift field $B_{S}$ of
16 tesla was obtained by fitting the linear region of $B_{c2}$ (see Fig. 5).

\begin{figure}
\centering \includegraphics[width=8cm,clip]{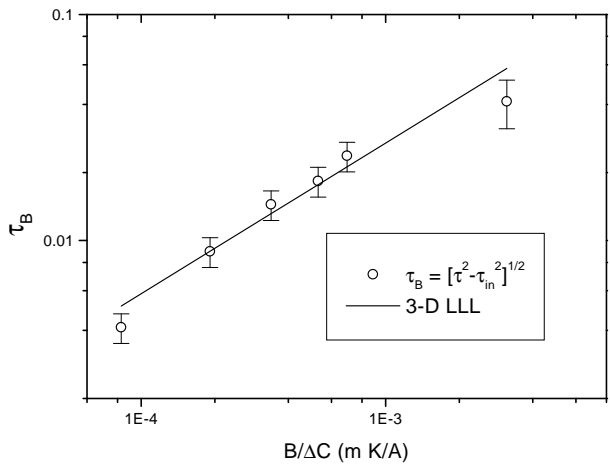}
\caption{ Field-broadened dimensionless parameter $\protect\tau _{B}$ is
plotted against $B/\Delta C$ on a log-log scale. The parameter was obtained
by using $\protect\tau _{B}=(\protect\tau^{2}-\protect\tau _{in}^{2})^{1/2}$
and $\Delta C$ is a mean-field discontinuity at T$_{c}$. The solid line
represents a lowest-Landau-level (LLL) fluctuation: $\protect\tau _{B}=%
\protect\beta (B/\Delta C)^{1/\protect\alpha }$ with $\protect\alpha =3/2$
and $\protect\beta =2.7$ $(A/m\cdot K)^{2/3}$.}
\label{figure6}
\end{figure}
Before making any further conclusions, we discuss some of the assumptions that we made in the analysis. In the zero-field analysis, sample inhomogeneity has been neglected. Boron 11 isotope Mg$^{11}$B$_{2}$ has a T$_{c}$ of 39 K while the excess specific heat extends well above 40 K. We might expect inhomogeneity effects to complicate the fluctuation analysis below the transition temperature but above T$_{c}$, where our analysis is concerned, the effect will be negligible. However, when nonzero field is applied, sample inhomogeneity must be considered because the analysis is of the field dependent behavior of the transition region. Sample inhomogeneity could produce an additional broadening through the Ginzburg-Landau parameter $\kappa $ , and hence H$_{c2}$. Information on H$_{c2}$ slopes at different parts of the sample with different T$_{c}$'s would be needed to account for the additional broadening correctly. To be more precise, the H$_{c2}$ slope in the T$_{c}$ = 39 K part of the sample and that in the, e.g., T$_{c}$ = 38 K part would be needed. In our in-field analysis, we assumed that the slopes of H$_{c2}$ at different parts of the sample are same or if they are different, the difference is small, which leads to field independent inhomogeneity effect. It is necessary to study high quality single crystals with different T$_{c}$'s to better understand sample inhomogeneity effects on transition-width broadening.

In summary, the zero-field specific heat was discussed in the context of BCS
theory plus 3D Gaussian fluctuations. The analysis indicates that Mg$^{11}$B$%
_{2}$ is a bulk superconductor and its coherence length is about 26 \AA .
In-field specific heat was treated in terms of lowest-Landau-level
fluctuations. That analysis requires that $\xi _{0}=20$ \AA . The in-field
analysis could be complicated due to the effect of anisotropy in a
polycrystalline sample. The anisotropy of $B_{c2}$ between ab-plane and
c-axis directions can lead to a field dependent broadening due to the T$_{c}$
distribution arising from the randomly oriented grains of the present
sample. This in turn leads to an underestimation of the G-L coherence
length. In order to understand the influence of anisotropy, we can assume
that the transition broadening arises solely from $B_{c2}$ anisotropy and
calculate the required anisotropy in our experimental temperature range. The
reported anisotropy value of $\sim $3 from single crystal measurements \cite%
{single-crystal1}$^{,}$\cite{single-crystal2}$^{,}$\cite{single-crystal3} is
much smaller than the ratio 6 required to explain the broadening. From this
consideration, we conclude that the anisotropy alone cannot explain the
whole broadening and therefore that fluctuation effects should be considered
in explaining the anomalous broadening.

This work at Urbana was supported by NSF DMR 99-72087. And the work at Pohang
was supported by the Ministry of Science and Technology of Korea through the
Creative Research Initiative Program.

\bibliography{mg11b2}

\end{document}